\documentclass[twocolumn]{aastex701}

\usepackage{amsmath}
\usepackage{booktabs}
\usepackage{threeparttable}
\usepackage{tabularx}

\usepackage{listings}

\begin{document}

\title{Dual Signatures of Bursty Star Formation in High-Redshift UV Luminosity Functions}

\author[orcid=0000-0003-4070-497X]{Guochao Sun}
\affiliation{CIERA and Department of Physics and Astronomy, Northwestern University, 1800 Sherman Ave, Evanston, IL 60201, USA}
\email[show]{guochao.sun@northwestern.edu}

\author[orcid=0000-0002-4900-6628]{Claude-Andr\'{e} Faucher-Gigu\`{e}re}
\affiliation{CIERA and Department of Physics and Astronomy, Northwestern University, 1800 Sherman Ave, Evanston, IL 60201, USA}
\email[]{cgiguere@northwestern.edu}

\author[orcid=0000-0002-0658-1243]{Steven R. Furlanetto}
\affiliation{Department of Physics and Astronomy, University of California, Los Angeles, CA 90095, USA}
\email[]{sfurlane@astro.ucla.edu}

\author[orcid=0000-0002-3950-9598]{Adam Lidz}
\affiliation{Department of Physics and Astronomy, University of Pennsylvania, 209 S. 33rd Street, Philadelphia, PA 19104, USA}
\email[]{alidz@sas.upenn.edu}

\begin{abstract}

The UV luminosity function (UVLF) encodes key information about galaxy formation. The slowly evolving bright-end UVLFs at $z\gtrsim10$ have made UV variability from bursty star formation a promising explanation, but the impact of such variability need not be restricted to the bright end. Motivated by ultra-deep measurements of the $z\simeq7$ UVLF, we investigate whether UV variability can provide a unified interpretation of the low-mass star formation efficiency (SFE) at $z\simeq7$ and the abundance of UV-bright galaxies at $z\gtrsim12$. Using abundance matching, we show that a steep faint-end UVLF at $z\simeq7$ admits a range of interpretations. For modest, mass-independent UV variability, the inferred median SFE starts flattening below $M_{\rm h} \sim10^{10.5} M_{\odot}$. If UV variability instead grows with decreasing halo mass---a possibility motivated by simulations and observations---the inferred median SFE is steeper and lower by a factor of about three at $\sim10^9 M_{\odot}$, consistent with stronger feedback suppression in low-mass halos. Calibrated to the common target $z\simeq7$ UVLF, these two scenarios diverge strongly when extrapolated to $12\lesssim z\lesssim17$ under redshift-independent UV variability and SFE prescriptions. Strongly mass-dependent $M_\mathrm{UV}$ scatter links the steep $z\simeq7$ faint-end slope and high $z\gtrsim12$ bright-end abundance as dual signatures of burstiness in similarly low-mass halos. Stronger variability makes bright galaxies contribute a larger fraction of the UV luminosity density, whereas weaker variability favors faint sources, despite their similar predicted reionization histories. At fixed $M_\mathrm{UV}$, UV variability broadens the halo mass distribution and lowers the clustering bias, which provides a useful diagnostic of burstiness for JWST and Roman. 

\end{abstract}



\section{Introduction} \label{sec:intro}

The ultraviolet luminosity function (UVLF) is one of the most important summary statistics of galaxy populations. It encodes key physical information about the halo mass function, the baryonic accretion rate, the efficiency with which accreted gas is converted into stars, and attenuation by interstellar dust. In particular, since UV emission traces the formation of short-lived massive stars in galaxies, interpretations of the UVLF are also affected by the recent star formation histories (SFHs) of galaxies, which are found to be highly time-variable at high redshift in both theory \citep[e.g.,][]{CAFG2018,FurlanettoMirocha2022,Sun2023a,Hopkins2023} and observations \citep[e.g.,][]{Dressler2024,Endsley2025,Munoz2026}. Therefore, UVLF measurements do not merely probe the median relation between the star formation rate (SFR) and halo mass, but also the scatter around it, which can be associated with both the time variability of individual halos and halo-to-halo variations \citep{Sun2023b,Samuel2026}. 

Observations of galaxies at cosmic dawn made by the James Webb Space Telescope (JWST) have highlighted the importance of this scatter. At $z\gtrsim12$, recent studies have reported a relatively high abundance of UV-bright galaxies and a slowly evolving bright end of the UVLF \citep[e.g.,][]{Donnan2024,Finkelstein2024,Harikane2023,Robertson2024,PerezGonzalez2025,Whitler2025}. This has motivated several possible explanations, including enhanced star formation efficiencies (SFEs) due to inefficient feedback \citep[e.g.,][]{Dekel2023}, a top-heavy stellar initial mass function \citep[IMF; e.g.,][]{Inayoshi2022}, dust-free stellar populations \citep[e.g.,][]{Ferrara2023}, hidden AGN contributions \citep[e.g.,][]{Hegde2024}, and departures from standard cosmology \citep[e.g.,][]{Shen2024}. Another plausible explanation is that early galaxies exhibit strongly bursty SFHs \citep[e.g.,][]{Mirocha2023,Shen2023,Sun2023b,Kravtsov2024}, in which case the bright-end UVLF is boosted by the Eddington bias associated with upward fluctuations in UV luminosity induced by recent starbursts, without requiring uniformly elevated SFEs. The boost is particularly strong at high redshift when the halo mass function is steep and dominated by the low-mass halos, whose SFHs remain highly bursty before the gravitational potential becomes sufficiently deep \citep[e.g.,][]{Stern2021,Gurvich2023,Hopkins2023}. 

While the impact of bursty star formation on the UVLF is more often discussed in the context of the bright end at very high redshifts, the same physics can in fact affect the faint end as well at somewhat lower redshift. With the aid of strong gravitational lensing, recent measurements by the GLIMPSE survey have pushed the $z \simeq 7$ UVLF to magnitudes as faint as $M_\mathrm{UV} \sim -13$ \citep{Atek2025,Atek2026}, approaching a regime unexplored by previous strong lensing analyses \citep[e.g.,][]{Ishigaki2018} and sensitive to star formation in extremely low-mass halos. At face value, the steep faint-end slope observed requires small halos to remain UV-bright. For a tight luminosity--halo mass relation, the steepness can be reproduced if low-mass halos sustain sufficiently high median SFEs. Instead, if UV variability grows substantially toward lower masses, increased fluctuations from the numerous intrinsically faint galaxies can enhance the faint-end abundance. The same steep faint-end UVLF can therefore be consistent with a lower and more steeply declining median SFE.

This is a crucial distinction for understanding feedback regulation in early galaxies. Analytic models and simulations commonly predict that stellar feedback suppresses star formation more strongly in lower-mass halos, leading to an SFE that increases with halo mass \citep[e.g.,][]{Furlanetto2017,Ma2018,Tacchella2018,Behroozi2019,Feldmann2025}. A substantially flattened median SFE could indicate weaker feedback suppression in low-mass halos than in models with a steeper SFE. Alternatively, substantial UV variability can reconcile a steep faint-end UVLF with a lower and steeper median SFE, consistent with stronger feedback suppression in low-mass halos. The key point is therefore that the faint-end slope is not uniquely determined by the median SFE, but instead reflects the convolution of the median galaxy–halo connection with UV variability, whose importance has been revealed by a number of earlier studies \citep[e.g.,][]{Ren2019,Shen2023,Sun2023b,Gelli2024}. 

In this paper, we show how the steep faint-end UVLF at $z\simeq7$ and the elevated bright-end UVLF at $z\gtrsim12$ can arise as dual signatures of burstiness in halos of similar low masses. We use the abundance matching technique to infer the median SFE for a family of UV variability prescriptions calibrated to produce the common target UVLF at $z\simeq7$ measured by GLIMPSE. We then extrapolate to higher redshifts under the simplified yet physically motivated assumption that both the median SFE and UV scatter are redshift-independent functions of halo mass. Our fiducial bursty model predicts substantially higher bright-end UVLFs over $10 \lesssim z \lesssim 17$ than the non-bursty model with a constant $0.5$\,mag scatter and remains broadly consistent with existing JWST observations. We also examine the resulting host halo distributions and galaxy clustering signals, including predictions for the recently launched Roman Space Telescope, and show how different burstiness scenarios yield different contributions from bright and faint galaxies to the UV luminosity density driving reionization. 

Throughout, we adopt cosmological parameters $\Omega_{\rm m} = 0.27$, $\Omega_{\Lambda} = 0.73$, $\Omega_{\rm b} = 0.045$, $h = 0.7$, $\sigma_8 = 0.8$, and $n_{\rm s} = 0.96$, consistent with the choices made in \citet{Trac2015}, and quote all magnitudes in the AB system.


\section{Models} \label{sec:models}

In this section, we present the empirical framework used to interpret high-$z$ UVLF measurements. The goal is to infer how the median SFE depends on halo mass under different assumptions about UV variability, and then to examine how these empirically inferred solutions behave when extrapolated to earlier cosmic epochs under simplified assumptions of redshift independence. We describe the halo growth prescription, the abundance matching framework, and the modeling of reionization and galaxy clustering. 

\subsection{Halo Evolution}

We anchor our model of the galaxy population to the evolution of halo mass $M_\mathrm{h}$ across redshift $z$. We describe the halo number statistics using the \citet{Trac2015} mass functions, $d n / d M_\mathrm{h}$, and estimate the halo mass growth rate, $\dot{M}_\mathrm{h}(M_\mathrm{h},z)$, by evolving halos at fixed cumulative number density through the abundance matching technique, namely
\begin{equation}
\int^{\infty}_{M_\mathrm{h,1}} \frac{d n (M_\mathrm{h}, z_1)}{d M_\mathrm{h}} d M_\mathrm{h} = \int^{\infty}_{M_\mathrm{h,2}} \frac{d n (M_\mathrm{h}, z_2)}{d M_\mathrm{h}} d M_\mathrm{h},
\end{equation}
which provides a self-consistent mapping between halo abundance and halo growth that compares reasonably well to results from numerical simulations as demonstrated in \citet{Furlanetto2017}. The baryonic mass growth rate is then
$\dot{M}_\mathrm{b} = f_\mathrm{b} \dot{M}_\mathrm{h}$, where $f_\mathrm{b} = \Omega_\mathrm{b}/\Omega_\mathrm{m}$. We assume that high-$z$ galaxies mainly grow by effectively smooth accretion of fresh gas onto their host halos and focus on their in-situ star formation, with major mergers making a subdominant contribution to early galaxy growth \citep[see e.g.,][]{Genel2010,Romano-Diaz2014,Angles-Alcazar2017}. This simple prescription is sufficient for our purpose, which is to compare how different assumptions about UV variability alter the median SFE inferred from the same target UVLF. 

\subsection{Empirical Abundance Matching Analysis} \label{sec:models:abundance_matching}

To infer the median UV luminosity--halo mass relation required to reproduce a target UVLF, we forward model the (dust-corrected) UVLF, $\Phi(M_\mathrm{UV})$, as a convolution of $d n / d M_\mathrm{h}$ with the conditional probability distribution of the UV magnitude $M_\mathrm{UV}$ at a given halo mass $M_\mathrm{h}$
\begin{equation}
\Phi(M_\mathrm{UV}) = \int d M_\mathrm{h} \frac{d n}{d M_\mathrm{h}} P(M_\mathrm{UV}|M_\mathrm{h}), 
\end{equation}
where the integration starts from $M_\mathrm{h,min}=10^8\,M_{\odot}$ roughly corresponding to the atomic cooling limit and we assume $P(M_\mathrm{UV}|M_\mathrm{h})$ to be a Gaussian distribution centered on a median $M_\mathrm{UV}$--$M_\mathrm{h}$ relation with a generally mass-dependent scatter $\sigma_\mathrm{UV}(M_\mathrm{h})$. For each prescribed $\sigma_\mathrm{UV}(M_\mathrm{h})$, we infer a regularized median $M_\mathrm{UV}$--$M_\mathrm{h}$ relation by matching the predicted cumulative UVLF to a common target $n(<M_\mathrm{UV})$.

\begin{figure*}
    \centering
    \includegraphics[width=0.9\textwidth]{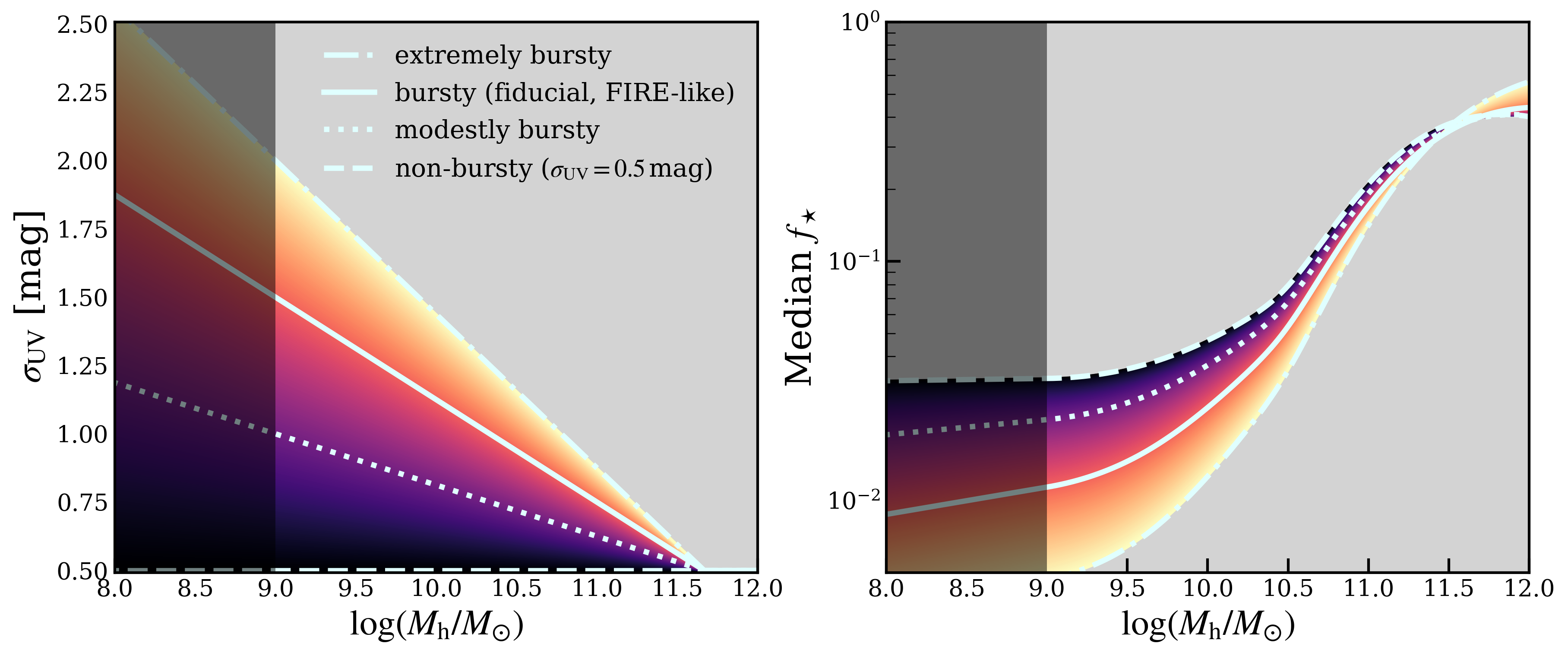}
    \caption{Flexible UV scatter prescriptions and the corresponding median SFEs inferred from abundance matching of the common target UVLF at $z \simeq 7$ measured by GLIMPSE \citep{Atek2026}. Left: A continuous family of physically motivated UV scatter parameterizations can be assumed to infer the SFE from the UVLF, as indicated by the color coding (each color representing a different model). Four specific model variations are highlighted. The fiducial \emph{bursty} model (solid line) assumes a strong mass dependence motivated by FIRE-2 simulations \citep{Sun2023b} and joint observations of H$\alpha$ and UV emission \citep{Munoz2026}. The \emph{extremely bursty} (dash-dotted line) and \emph{modestly bursty} (dotted line) variations extend this model to higher and lower burstiness, respectively. Finally, the \emph{non-bursty} model (dashed line) with $\sigma_\mathrm{UV}=0.5$\,mag represents a baseline value motivated by the scatter of halo mass accretion rate. Right: Median SFE as a function of mass required to reproduce the same dust-corrected $z \simeq 7$ UVLF for each color-coded scatter prescription. In the non-bursty case, the steep faint-end UVLF requires a substantially flattened SFE in $\lesssim10^{10.5}\,M_\odot$ halos. As the UV scatter grows toward lower halo masses, the inferred median SFE becomes progressively steeper, with the fiducial SFE being approximately three times lower at $10^9\,M_{\odot}$. Shading marks the low-mass regime where the mass-dependence of $\sigma_\mathrm{UV}$ is highly uncertain and the median SFE adopts a power-law extrapolation from the inferred slope at $M_\mathrm{h}=10^9\,M_{\odot}$.}
    \label{fig:sfe_am}
\end{figure*}

As illustrated in the left panel of Figure~\ref{fig:sfe_am}, for our baseline analysis, we consider a continuous family of UV scatter prescriptions parameterized by a variable mass slope, an anchor mass of $\sim10^{11.7}M_{\odot}$, and a minimum value of $0.5$\,mag at high halo masses motivated by the level of scatter in the halo mass accretion rate \citep{Mirocha2021}. Among the parameterized models, we specifically consider a \emph{non-bursty} model with a fixed mass-independent scatter of 0.5\,mag and a fiducial \emph{bursty} model with $\sigma_{\rm UV}$ reaching 1.5\,mag at $10^9 M_{\odot}$, motivated by previous theoretical predictions of bursty star formation in low-mass galaxies and by observational constraints from joint analyses of H$\alpha$ and UV emission \citep{Sun2023b,Gelli2024,Munoz2026}. Two additional reference models, \emph{modestly bursty} and \emph{extremely bursty}, with $\sigma_{\rm UV}(10^9 M_{\odot}) = 1.0$ and 2.0\,mag, respectively, are considered. 

We then study the mass dependence of the SFE implied by the median $M_\mathrm{UV}$--$M_\mathrm{h}$ relation inferred. A given inferred $M_\mathrm{UV}$, or $L_{\mathrm{UV}}$, can be related to the SFR by
\begin{equation}
\dot{M}_{\star} = \mathcal{K}_\mathrm{UV} L_\mathrm{UV}, 
\end{equation}
where the conversion factor $\mathcal{K}_\mathrm{UV}$ is taken to be $1.15 \times 10^{-28}\,M_{\odot}\,\mathrm{yr^{-1}/erg\,s^{-1}\,Hz^{-1}}$ \citep{MadauDickinson2014}, although in practice it depends on the SFH, metallicity, and initial mass function. The SFE inferred from the target UVLF is then
\begin{equation}
f_{\star}(M_\mathrm{h}) = \dot{M}_{\star} / \dot{M}_\mathrm{b} = \dot{M}_{\star} / (f_\mathrm{b}\dot{M}_\mathrm{h}). 
\label{eq:sfe}
\end{equation}
The shape of $f_{\star}(M_\mathrm{h})$ encodes the physical information contained in the UVLF about how efficiently the baryonic mass supply is turned into stars and therefore the strength of feedback regulation \citep{Dutton2009,Furlanetto2017}. 

\subsection{High-$z$ Extrapolation and EoR Implications}

To connect the $z\simeq7$ faint-end UVLF to the bright end at $z\gtrsim12$, we extrapolate the abundance matching results to higher redshifts under the assumption that neither the inferred median SFE nor $\sigma_{\rm UV}(M_\mathrm{h})$ evolves with redshift. While simplified, this assumption serves as a baseline motivated by theoretical predictions according to which much of the observed evolution follows that of the halo abundance and accretion rates \citep{Feldmann2025}. For a given model obtained from the $z\simeq7$ UVLF, we compute the SFR and $M_{\rm UV}$ of halos at the higher redshift and then convolve the resulting median $M_{\rm UV}$–$M_{\rm h}$ relation with the corresponding $\sigma_{\rm UV}(M_{\rm h})$. This extrapolation is not intended to be an exact physical model of early galaxy evolution. Rather, it provides a flexible way to explore whether the different $z\simeq7$ interpretations lead to qualitatively different and distinguishable predictions at much higher redshift. Dust attenuation is assumed to be negligible at $z\gtrsim10$. 

Based on the extrapolated UVLFs, we also compute the ionizing photon emissivity to examine the reionization histories implied. We first integrate the intrinsic (i.e., dust-corrected) UV luminosity density down to the minimum mass of star-forming halos $M_\mathrm{h,min}$,
\begin{equation}
\rho_{\rm UV}^{\rm int} = \int
dM_{\rm h} \frac{d n}{d M_{\rm h}} \int d L_{\rm UV} L_{\rm UV} P(L_{\rm UV}|M_{\rm h}).
\end{equation}
We then estimate the comoving ionizing emissivity as
\begin{equation}
\dot{n}_{\rm ion} = f_{\rm esc}\xi_{\rm ion}\rho_{\rm UV}^{\rm int},
\end{equation}
where $f_{\rm esc}$ is the escape fraction of ionizing photons and $\xi_{\rm ion}$ is the ionizing photon production efficiency defined relative to the intrinsic UV luminosity after dust correction. We adopt $f_{\rm esc}=0.1$ and $\log (\xi_{\rm ion}/{\rm Hz\,erg^{-1}})=25.2$. We translate the ionizing emissivity into a volume-averaged reionization history by evolving the ionized filling factor $Q_{\rm HII}$ according to
\begin{equation}
\frac{dQ_{\rm HII}}{dt}
=\frac{\dot n_{\rm ion}}{\bar n_{\rm H,0}}
-\frac{Q_{\rm HII}}{t_{\rm rec}},
\end{equation}
where
\begin{equation}
t_{\rm rec}^{-1}
=C \alpha_{\rm B}
\left(1+\frac{Y_{\rm p}}{4X_{\rm p}}\right)
\bar n_{\rm H,0}(1+z)^3,
\end{equation}
$X_{\rm p}=0.76$ and $Y_{\rm p}=0.24$ are hydrogen and helium mass fractions, and $\bar n_{\rm H,0}$ is the comoving hydrogen number density. We assume the commonly adopted clumping factor $C=3$ for the ionized IGM \citep[but see, e.g.,][for higher values implied by recent observations at $z\lesssim6$]{Davies2024} and $\alpha_{\rm B}=2.6\times10^{-13}\,{\rm cm^3\,s^{-1}}$ for the case-B recombination coefficient at $10^4\,$K. 

\subsection{Galaxy Clustering Analysis}

UV variability changes not only the abundance of galaxies at a given UV magnitude, but also the distribution of their host halo masses and hence their clustering \citep[e.g.,][]{Munoz2023,Gelli2024,Sun2025,ChakrabortyChoudhury2026}. Following \citet{Munoz2023} and \citet{Gelli2024}, we define the galaxy number density and number-weighted effective linear bias in a given range of $M_\mathrm{UV}$ to be
\begin{equation}
n
=
\int dM_{\rm h}\,
\frac{dn}{dM_{\rm h}}\,
\int
d M_{\rm UV} P(M_{\rm UV} | M_{\rm h}),
\end{equation}
and
\begin{equation}
\!b_{{\rm eff}}\!
=\!
n^{-1}
\!\!\int\!
dM_{\rm h}\,
\frac{dn}{dM_{\rm h}}\,
b(M_{\rm h})\,
\!\!\int\!
d M_{\rm UV} P(M_{\rm UV} | M_{\rm h}),
\end{equation}
respectively, where $b(M_\mathrm{h})$ is the halo bias defined for an overdensity of 200 relative to the mean matter density \citep{Tinker2010}. We evaluate $b_\mathrm{eff}$ for both cumulative selection brighter than a given magnitude and magnitude bins of width 1\,mag. Larger UV variability allows more numerous, less-biased low-mass halos to enter a given
magnitude bin, thereby lowering
$b_{\rm eff}$.

The constraining power on $b_\mathrm{eff}$ from galaxy clustering measurements can be estimated by modeling the angular clustering power spectrum using the Limber approximation \citep{Limber1953}
\begin{equation}
C_{\ell}
=
b_{{\rm eff}}^2
\int dz\,
\frac{H(z)}{c\chi^2(z)}
W_z^2(z)
P_{\rm m}\left(\frac{\ell+1/2}{\chi(z)},z\right),
\end{equation}
where $W_z$ is a normalized top-hat redshift window function, $\chi$ is the comoving radial distance, and $P_\mathrm{m}$ is the matter power spectrum. Assuming Poisson
sampling, the shot noise contribution to the angular power spectrum is
\begin{equation}
N_{\ell}
=
\frac{1}{\Sigma},
\end{equation}
where $\Sigma$ is the predicted galaxy surface number density per steradian. The variance for a bandpower of sky fraction $f_\mathrm{sky}$ and multipole bin width $\Delta\ell$ is then given by the Knox formula \citep{Knox1995}
\begin{equation}
\left( \delta C_{\ell} \right)^2
=
\frac{
\left(C_{\ell}+N_{\ell}\right)^2
}{
(\ell+1/2)\Delta\ell f_{\rm sky}
}.
\end{equation}
We use the Fisher matrix formalism to estimate the uncertainty on the bias parameter, namely
\begin{equation}
F_{bb}
=
\sum_{\ell}
\frac{
\left(
\partial C_{\ell}/
\partial b_{{\rm eff}}
\right)^2
}{
\left(\delta C_{\ell}\right)^2
},
\end{equation}
which gives a theoretical lower bound (the Cramér-Rao bound) on the uncertainty of $b_\mathrm{eff}$
\begin{equation}
\sigma(b_{{\rm eff}}) = F_{bb}^{-1/2}. 
\end{equation}

To forecast the constraints on $b_\mathrm{eff}$ that Roman can achieve, we consider the 2415\,deg$^2$ Medium-Tier and the 19.2\,deg$^2$ Deep-Tier observations of the Roman High-Latitude Wide-Area Survey (HLWAS) that have the appropriate filter combinations for dropout selections at the redshifts of interest. We further assume redshift bins of half width $\Delta z = 0.5$, and $100 \leq \ell \leq 10000$ with $\Delta\ell=100$. Our predicted uncertainties assume a fixed cosmology and include sample variance and Poisson noise only. Since realistic selection functions, non-Gaussian covariance, or other observational systematics are not taken into account, they should be understood as optimistic estimates.


\section{Results} \label{sec:results}

\subsection{SFE Constrained by Abundance Matching}

The abundance matching technique \citep{ValeOstriker2004} has been widely used to constrain the SFE of high-$z$ galaxies from HST and JWST observations \citep[e.g.,][]{Mason2015,SF2016,Tacchella2018,SippleLidz2024,Yung2025}. With the aid of strong lensing, these constraints can now be extended to substantially lower masses. In Figure~\ref{fig:sfe_am}, we show the median SFE inferred from the dust-corrected UVLF at $z\simeq7$ measured by the GLIMPSE survey, using the abundance matching technique described in Section~\ref{sec:models:abundance_matching}. We stress that the UV scatter is allowed to vary with halo mass in a continuous family of models as shown in the left panel (each corresponding to a different color). 

For modest, mass-independent $\sigma_{\rm UV}$, the $M_{\rm UV}$--$M_{\rm h}$ relation is tight. Reproducing the steep faint-end slope in this case requires the median SFE to substantially flatten below $M_{\rm h}\sim10^{10.5} M_\odot$, so that comparatively efficient star formation is maintained in low-mass halos\footnote{On the contrary, a steeper SFE in this case would map a fixed range of $M_{\rm h}$ onto a wider range of $M_{\rm UV}$ and produce a shallower faint-end UVLF.}. This implies that stellar feedback becomes less efficient at suppressing star formation than in models with more steeply declining low-mass SFEs. 

The interpretation becomes qualitatively different if $\sigma_{\rm UV}$ instead grows toward lower halo masses. As $\sigma_{\rm UV}$ increases more strongly toward lower halo masses, the inferred low-mass median SFE becomes progressively steeper. In the fiducial \emph{bursty} model, it is approximately three times lower at \(10^9\,M_\odot\) than in the \emph{non-bursty} model. The same target UVLF therefore permits vastly different low-mass SFE slopes and normalizations. Because $\sigma_{\rm UV}$ is allowed to increase steeply toward lower masses, the abundance matching results can be sensitive to the assumed minimum halo mass, $M_\mathrm{h,min}$, although we verify that our inferred qualitative trends are robust to order-unity variations in $M_\mathrm{h,min}$, especially at $M_\mathrm{h} \gtrsim 10^9 M_{\odot}$. Given the numerical sensitivity of the inferred low-mass SFE and the lack of empirical constraints or predictions, we adopt a power-law extrapolation of the inferred SFE below $10^9 M_{\odot}$, preserving the local slope, and use it for all subsequent calculations.

\begin{figure*}
    \centering
    \includegraphics[width=0.9\textwidth]{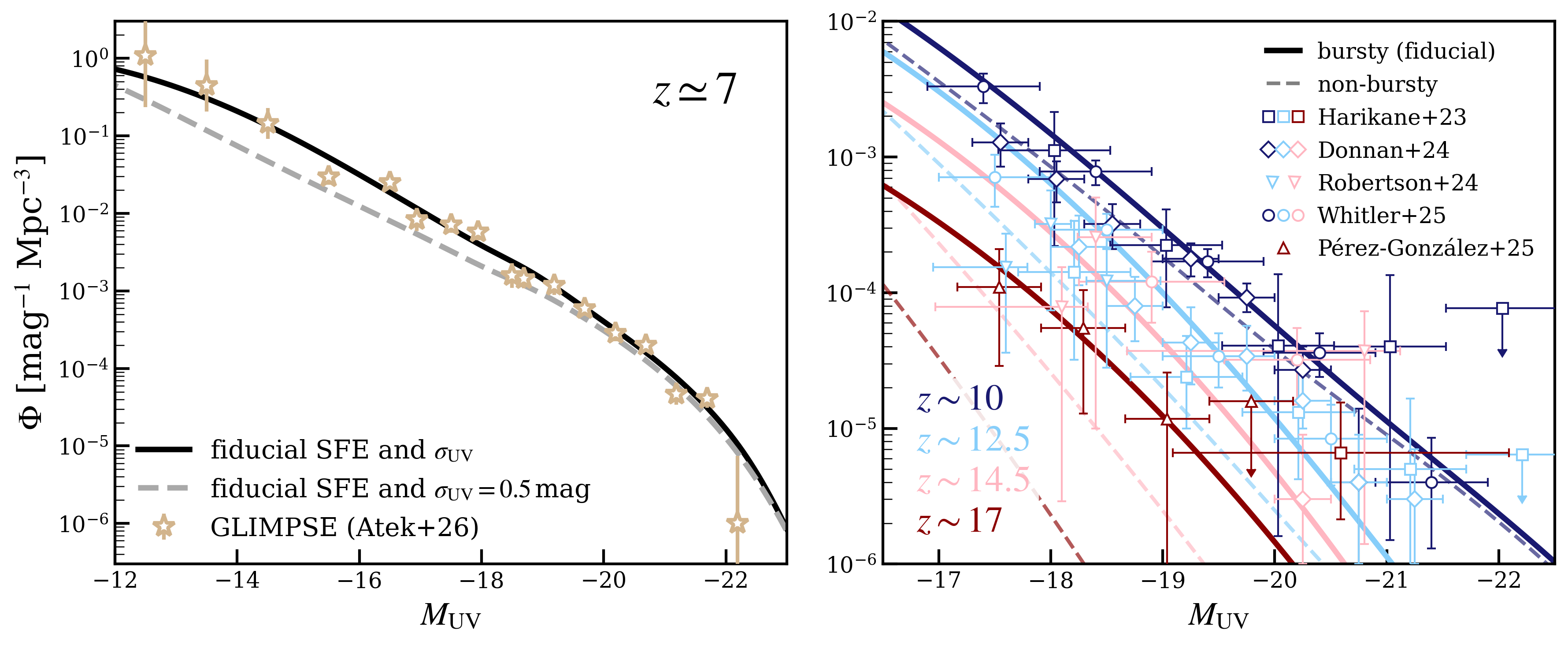}
    \caption{Left: The UVLF at $z\simeq7$ measured by the GLIMPSE survey \citep{Atek2026}, compared with predictions of the fiducial \emph{bursty} model that adopts the strongly mass-dependent $\sigma_{\rm UV}$ (black solid curve). The gray dashed curve keeps the same median SFE but adopts the same $\sigma_{\rm UV}=0.5$\,mag as in the \emph{non-bursty} model without refitting. This controlled comparison demonstrates the enhancement of faint-end abundance produced by a mass-dependent $\sigma_{\rm UV}$. Right: Comparison of $z\gtrsim10$ UVLFs predicted by extrapolating the abundance matching results at $z\simeq7$ to higher redshifts. Both UV variability and the median SFE are held fixed with redshift. The solid curves represent the fiducial \emph{bursty} model, whereas the dashed curves represent the \emph{non-bursty} model with $\sigma_\mathrm{UV}=0.5\,$mag. Although both models are calibrated to the common $z\simeq7$ UVLF by construction, they diverge strongly when extrapolated to $z > 10$. Extrapolations of the \emph{bursty} model remain broadly comparable to current JWST constraints on the abundance of UV-bright galaxies \citep{Harikane2023,Donnan2024,Robertson2024,Whitler2025,PerezGonzalez2025}, whereas the \emph{non-bursty} model increasingly underpredicts the observed number densities toward higher redshifts.}
    \label{fig:uvlf}
\end{figure*}

\subsection{How Burstiness Shapes the Faint-end UVLF}

To more explicitly show how UV variability shapes the $z\simeq7$ UVLF, especially its faint-end slope, we perform a controlled experiment where the median SFE as a function of halo mass is held fixed at the inferred result of the fiducial bursty case, while $\sigma_{\rm UV}$ is set to the value in the non-bursty case (0.5\,mag). This isolates the effect of varying $\sigma_{\rm UV}$'s mass dependence from the compensating adjustment of the median SFE relation as part of the abundance matching analysis. 

As shown by the comparison between the solid and dashed curves in the left panel of Figure~\ref{fig:uvlf}, when the median SFE is held fixed, removing the steep increase of $\sigma_{\rm UV}$ toward lower halo masses flattens the faint-end UVLF by a significant amount discernible by the GLIMPSE measurements. This supports the interpretation of Figure~\ref{fig:sfe_am}. The fact that removing the mass dependence of $\sigma_{\rm UV}$ at fixed median SFE reduces the predicted faint-end abundance suggests that luminosity fluctuations are responsible for a substantial fraction of the galaxies in this regime. Reproducing the target UVLF with a small $\sigma_{\rm UV}$ would thus require an increased low-mass median SFE to compensate for this deficit.

\subsection{Connection to Ultrahigh-Redshift UVLFs} \label{sec:ultrahighz}

\begin{figure}[!ht]
    \centering
    \includegraphics[width=0.95\columnwidth]{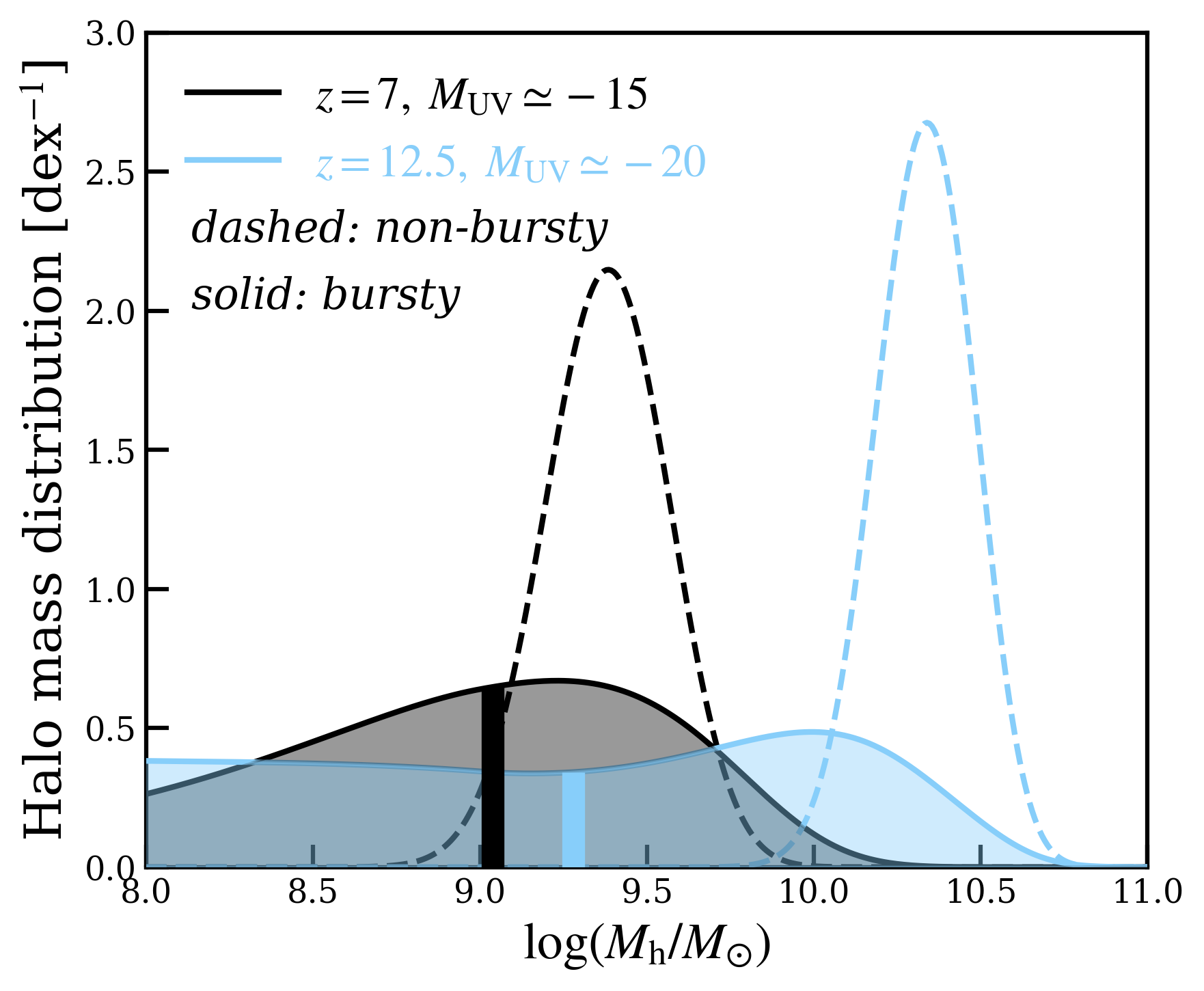}
    \caption{Distribution of halo masses contributing to the faint-end UVLF at $z=7$ (black) and the bright-end UVLF at $z=12.5$ (blue), respectively, in 1\,mag-wide bins. Shaded regions with solid edges represent the \emph{bursty} model, whereas unshaded regions with dashed edges represent the \emph{non-bursty} model. Each distribution is normalized to unity. Growing UV variability broadens the distribution toward lower masses. In our fiducial \emph{bursty} case, the distributions overlap substantially and have median masses (thick vertical bars) that differ by less than a factor of two, in contrast to the order-of-magnitude difference in the \emph{non-bursty} case. Halos of similar low masses thus contribute to the UV-faint and UV-bright populations at $z=7$ and $z=12.5$, respectively.}
    \label{fig:dual}
\end{figure}

To assess whether UVLF measurements at ultrahigh redshifts can help distinguish between these interpretations of the $z \simeq 7$ UVLF, we extrapolate the abundance matching solutions to $z\gtrsim10$ under the simplified assumption that both the median SFE and $\sigma_{\rm UV}$ are non-evolving with redshift (see Section~\ref{sec:summary} for discussion of this assumption). In the right panel of Figure~\ref{fig:uvlf}, we compare the resulting UVLF predictions from extrapolation---both between the \emph{bursty} vs. \emph{non-bursty} scenarios and against recent JWST observations at $10\lesssim z\lesssim17$ \citep{Harikane2023,Donnan2024,Robertson2024,PerezGonzalez2025, Whitler2025}.

Although the \emph{bursty} and \emph{non-bursty} models reproduce the same $z\simeq7$ UVLF by construction, they diverge significantly when extrapolated to higher redshifts. The \emph{bursty} model with strongly mass-dependent $\sigma_{\rm UV}$ predicts relatively modest evolution in the bright-end UVLF: compared against its value at $z=7$, $\Phi(M_{\rm UV}=-20)$ decreases by roughly 0.8\,dex and 1.9\,dex by $z\sim10$ and 15, respectively. Its predictions are therefore broadly comparable to the observed abundance of photometric galaxy candidates out to $z\sim17$. By contrast, the \emph{non-bursty} model with $\sigma_{\rm UV}=0.5$ mag falls significantly short at $z > 12$, with discrepancies exceeding an order of magnitude at the highest redshifts. Under our simplified assumption of redshift-independent median SFE and $\sigma_{\rm UV}$, this suggests that the high number densities of bright $z\gtrsim12$ galaxies cannot be reconciled simply by invoking a flattened low-mass SFE. Instead, they may reflect the same mass-dependent UV variability that provides an alternative interpretation of the steep faint-end UVLF at $z\simeq7$. We note that qualitatively similar analyses that link the high-$z$ bright-end UVLF to mass-dependent UV variability can be found in \citet{Gelli2024} who adopt a fiducial SFE relation calibrated at $z\sim5$ and \citet{Munoz2026} who jointly constrain the SFE and burstiness using H$\alpha$/UV ratios and UVLFs. Quantitatively, UVLFs predicted by our \emph{bursty} model over $10 \lesssim z \lesssim 17$ are broadly aligned with those from \citet{Munoz2026} but several times higher than those from \citet{Gelli2024} at $z > 12$.

\begin{figure*}
    \centering
    \includegraphics[width=0.9\textwidth]{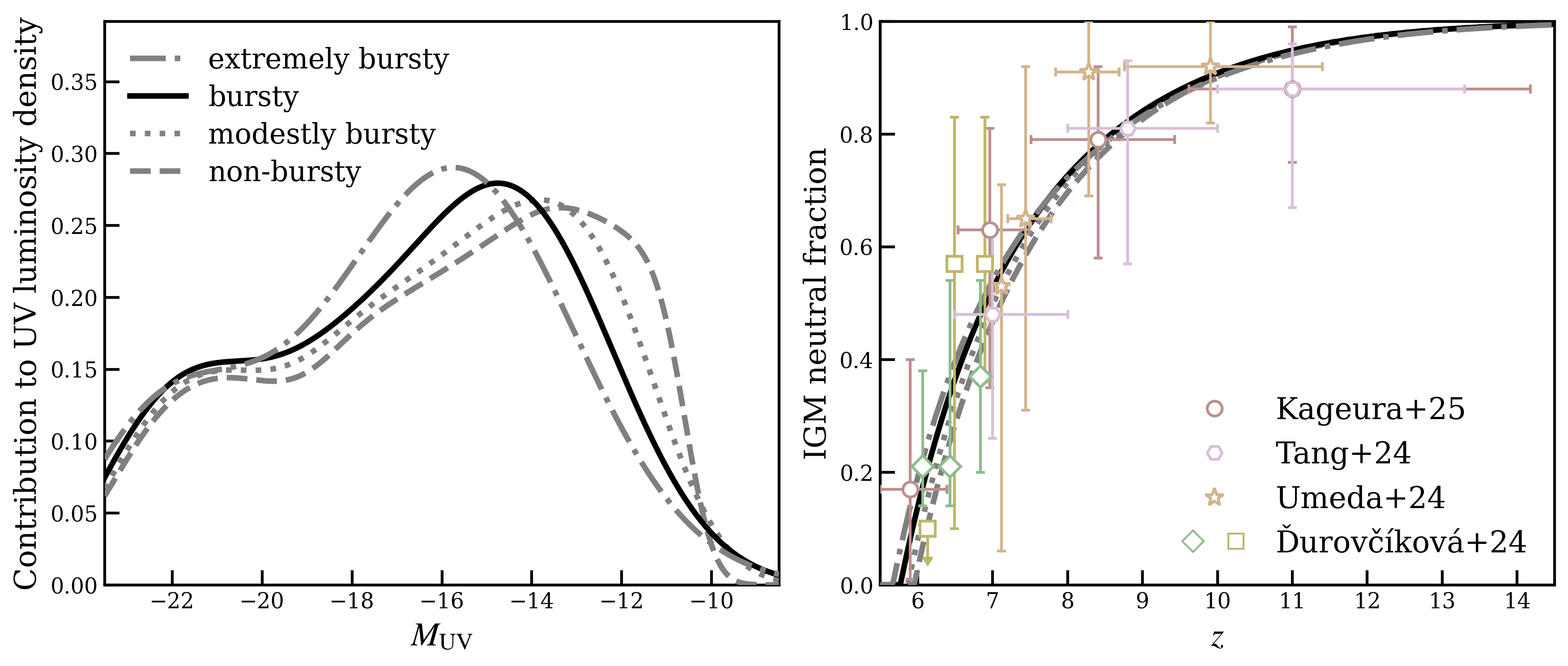}
    \caption{Left: Normalized fractional contribution to the UV luminosity density predicted by the four reference models at $z=7$, near the midpoint of reionization. Stronger variability shifts a larger fraction of the luminosity density to bright galaxies. Right: Redshift evolution of the IGM neutral fraction predicted by integrating the four reference models down to $10^8\,M_{\odot}$, assuming the ionizing escape fraction $f_\mathrm{esc}=0.1$, ionizing efficiency $\log (\xi_{\rm ion}/{\rm Hz\,erg^{-1}})=25.2$, and IGM clumping factor $C = 3$. Several representative observational constraints based on the Ly$\alpha$ luminosity function evolution \citep{Kageura2025}, Ly$\alpha$ emission equivalent widths \citep{Tang2024}, Ly$\alpha$ damping wing profiles \citep{Umeda2024}, and quasar absorption spectra \citep{Durovcikova2024} are shown for comparison. Despite the less rapidly declining bright-end UVLF predicted, more bursty models have a smaller contribution from faint sources and a lower total ionizing emissivity responsible for their slightly delayed reionization histories. }
    \label{fig:eor}
\end{figure*}

The possible connection between the steep faint-end UVLF at $z\simeq7$ and the elevated bright-end UVLF at $z\gtrsim12$ can be made more explicit by examining the halo masses contributing to $\Phi(M_{\rm UV})$ in a given UV magnitude regime. In Figure~\ref{fig:dual}, we show the normalized contribution from different halo masses to opposite regimes of the UVLFs at $z=7$ and $z=12.5$, contrasting our two reference models at the same time. The mass dependence of $\sigma_{\rm UV}$ in the \emph{bursty} model develops an extended tail in the distribution toward lower masses. Interestingly, the overlap between the halos contributing to the faint end at $z\simeq7$ (centered on $M_{\rm UV}=-15$) and the bright end at $z\simeq12.5$ (centered on $M_{\rm UV}=-20$) indicates that these two apparently distinct UVLF regimes probe signatures of similar populations of bursty, low-mass halos at different epochs. This is physically expected given the strong redshift dependence of the halo mass accretion rate, which increases roughly as $(1+z)^{2.5}$ at fixed halo mass \citep{Dekel2013,Furlanetto2017}, and substantial UV variability. The peaks of the two distributions need not coincide because the characteristic SFR at fixed halo mass evolves strongly with redshift. Nevertheless, the substantially overlapping distributions in the fiducial model have comparable median values, whereas the distributions in the \emph{non-bursty} case are much narrower and separated. This comparison demonstrates that the same characteristic halo population can contribute to genuinely separated regimes of the UVLF at two distinct cosmic epochs. 

It is noteworthy that models with a large constant $\sigma_{\rm UV}\sim1.5$--2\,mag can also produce high-$z$ bright-end UVLFs comparable to those predicted by our \emph{bursty} model (see e.g., earlier work by \citet{Shen2023} invoking UV variability to explain the JWST observations). However, these models imply shallower low-mass median SFEs. Among our examined prescriptions, the steeper low-mass SFE is therefore a more distinctive consequence of mass-dependent UV variability.

\subsection{Implications for Reionization}

\begin{figure*}[!ht]
    \centering
    \includegraphics[width=0.95\textwidth]{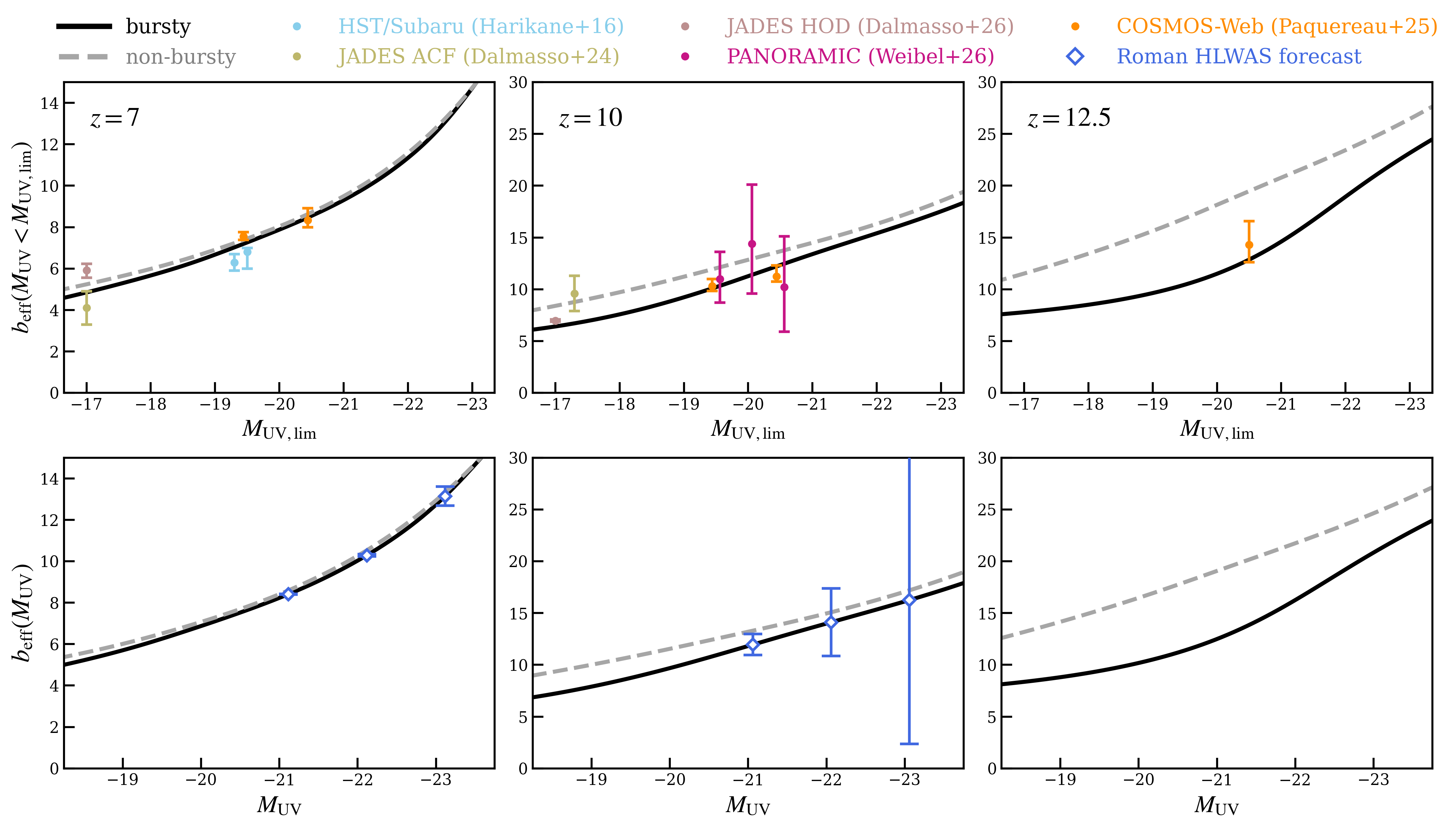}
    \caption{Top: Effective clustering bias, $b_\mathrm{eff}$, of galaxies brighter than a given limiting magnitude $M_\mathrm{UV,lim}$ at $z=7$ (left), 10 (middle), and 12.5 (right) predicted by the \emph{bursty} vs. \emph{non-bursty} models. Enhanced UV variability allows more numerous, lower-mass halos to populate bright luminosity bins, thereby lowering the predicted $b_\mathrm{eff}$ relative to the case with lower $\sigma_\mathrm{UV}$. For comparison, we show measurements from HST \citep{Harikane2016} and JWST observations \citep{Dalmasso2024,Paquereau2025,Dalmasso2026,Weibel2026}. Bottom: The same as the top panels but for galaxies in UV magnitude bins of width 1\,mag. Blue data points and error bars show the predicted $1\sigma$ constraints on $b_\mathrm{eff}$ for dropout samples selected from the Roman High-Latitude Wide-Area Survey (HLWAS). The $z=7$ predictions assume the 2415\,deg$^2$ Medium-Tier survey (aided by optical non-detections from Rubin/LSST) with $m_{\rm AB,lim}=26.4$, whereas the $z=10$ predictions assume the 19.2\,deg$^2$ Deep-Tier observations with $m_{\rm AB,lim}=27.0$. The nominal photometric depth and filter combinations of HLWAS do not give robust constraints at $z=12.5$.}
    \label{fig:bias}
\end{figure*}

In Figure~\ref{fig:eor}, we show how the four reference models distribute the UV luminosity density differently among galaxies of different $M_{\rm UV}$ and compare their predicted reionization histories with some representative observational constraints \citep{Durovcikova2024,Tang2024,Umeda2024,Kageura2025}. In the left panel, stronger variability moves the dominant contribution toward brighter magnitudes, whereas less bursty models receive a larger fractional contribution from fainter sources. Nevertheless, integrating the full luminosity distributions of halos above $M_{\mathrm{h,min}}$ produces similar reionization histories under the same assumptions for ionizing photon production, escape fraction, and IGM clumping factor, as suggested by the right panel. This is because differences in the predicted bright-end abundances (Figure~\ref{fig:uvlf}) are largely offset by differences in the faint populations associated with the separately abundance-matched median SFE relations. These relations yield similar mean UV luminosities in low-mass halos that increasingly dominate $\rho_{\rm UV}$ at higher redshifts and this similarity is preserved by the redshift-independent SFE and $\sigma_{\rm UV}$ assumed. Thus, a high bright-end UVLF does not necessarily imply earlier reionization. Although the exact reionization histories depend on the assumed model parameters, their broad agreement with observations provides a useful check on our extrapolations from the calibration at $z\simeq7$. 

\subsection{Galaxy Clustering and Roman Forecasts} \label{sec:clustering}

In Figure~\ref{fig:bias}, we show the effective clustering bias predicted by the \emph{bursty} and \emph{non-bursty} models for galaxies brighter than a given limiting magnitude $M_{\rm UV,lim}$ (top row) and in $M_{\rm UV}$ bins of width 1\,mag (bottom row). For comparison, several cumulative measurements of $b_\mathrm{eff}(M_\mathrm{UV}<M_\mathrm{UV,lim})$ from HST/Subaru \citep{Harikane2016} and JWST \citep{Dalmasso2024,Paquereau2025,Dalmasso2026,Weibel2026} are overplotted in the top panels, whereas predictions for the constraints Roman HLWAS will enable are shown in the bottom panels. Our \emph{bursty} model generally predicts a lower effective bias because enhanced UV variability allows the more numerous, less biased halos to enter a given luminosity bin. The difference is modest at $z=7$, where both models are matched to the same UVLF, but grows toward higher redshifts. We note that our analysis here differs in nature from \citet{Munoz2023}, who fit high-SFE and high-scatter solutions to the available UVLFs and used clustering to break the remaining degeneracy. We instead calibrate models with varying $\sigma_{\rm UV}(M_{\rm h})$ prescriptions using abundance matching at $z\simeq7$ and extrapolate the results to higher redshifts. Our clustering comparison therefore tests the extrapolated \(z\simeq7\) scenarios, which already predict different high-$z$ UVLFs as shown in Figure~\ref{fig:uvlf}, rather than breaking a degeneracy between enhanced UV variability and SFE to reproduce the same high-$z$ UVLFs as in \citet{Munoz2023}.  

Our Roman forecasts are tailored to the information available for photometric samples given the filter combinations and depths of HLWAS. Requiring two detections redward of the Ly$\alpha$ break, we consider F129/F158 for the Medium Tier at $z=7$ and F158/F184 for the Deep Tier at $z=10$, corresponding to adopted limiting magnitudes of $m_{\rm AB,lim}=26.4$ and 27.0, respectively. The enormous area (2415\,deg$^2$) of Medium-Tier HLWAS observations, aided by optical non-detections from Rubin/LSST for robust dropouts, yields tight constraints on $b_{\rm eff}$ at $z=7$. On the other hand, despite containing more filters, the 19.2\,deg$^2$ Deep-Tier observations have higher Poisson noise due to small source counts, which substantially weakens the clustering constraints at $z \gtrsim 10$. At $z\sim12.5$, requiring detections in both F184 and F213 limits the sample to the F213 depth of 25.9\,mag, which leaves too few galaxies for useful clustering constraints. Although observations with Roman's grism, which covers wavelengths up to 1.93\,$\mu$m, could enlarge the usable sample, quantifying the improvement requires dedicated sensitivity forecasts for grism spectroscopy, which we leave for future work. In Appendix~\ref{sec:appendix}, we present an additional diagnostic based on stellar masses implied by the inferred SFE. 


\section{Summary and Discussion} \label{sec:summary}

We have studied how UV variability associated with bursty star formation can manifest itself in high-$z$ galaxy UVLFs, with a particular focus on the possible physical connection between the steep faint-end slope at $z\simeq7$ and the high abundance of UV-bright galaxies at $z\gtrsim12$ revealed by recent JWST observations. Our main findings are summarized as follows:

\begin{itemize}

\item The steep faint-end UVLF at $z\simeq7$ from the GLIMPSE survey does not uniquely determine the median SFE of low-mass halos. Our abundance matching analysis shows that a mass-independent $\sigma_{\rm UV}=0.5\,$mag requires the median SFE to flatten substantially below $M_\mathrm{h} \sim 10^{10.5}M_{\odot}$. Scatter that grows toward lower halo masses instead permits a steeper median SFE, approximately three times lower at $10^9\,M_{\odot}$ in our fiducial \emph{bursty} model, consistent with stronger feedback suppression in low-mass halos (Figure~\ref{fig:sfe_am}). A controlled comparison at fixed median SFE further demonstrates that increased UV variability toward lower masses steepens the faint-end UVLF (Figure~\ref{fig:uvlf}).

\item Calibrated to the same $z \simeq 7$ UVLF by construction, our bursty and non-bursty models diverge strongly when extrapolated to earlier times under redshift-independent SFE and $\sigma_{\rm UV}$ prescriptions. The bursty model remains broadly comparable to current observations, whereas the non-bursty model increasingly underpredicts the galaxy abundance at $z \gtrsim 12$  (Figure~\ref{fig:uvlf}). In the fiducial bursty model, halo distributions for faint-end galaxies at $z=7$ and bright-end galaxies at $z=12.5$ overlap substantially, suggesting that these distinct UVLF regimes trace burstiness in similarly low-mass halos at different epochs (Figure~\ref{fig:dual}).

\item Stronger UV variability places a larger fraction of the UV luminosity density in bright galaxies, whereas weaker UV variability favors faint sources. However, integrating the full luminosity distributions of all star-forming halos in our reference models produces similar reionization histories under common assumptions for modeling the reionization (Figure~\ref{fig:eor}). Differences in the bright-end abundance are compensated by the faint populations due to separately calibrated median SFEs. 

\item Strong UV variability also broadens the galaxy--halo connection and decreases the effective clustering bias $b_{\rm eff}$ of galaxies. The contrast between our bursty and non-bursty models is modest at $z\simeq7$, but grows toward higher redshift (Figure~\ref{fig:bias}). Our forecasts for the Roman HLWAS indicate significant constraining power on $b_{\rm eff}$ from observations at $z\simeq7$ and 10, whereas increased Poisson noise due to smaller samples limits clustering measurements from photometric galaxies at $z \gtrsim 12$. 

\end{itemize}

Several caveats and limitations should be kept in mind when interpreting our results. Constraints on the extremely faint end of the $z\simeq7$ UVLF from the GLIMPSE survey rely on strong lensing magnification over small effective survey volumes. Although the steepness of the $z\simeq7$ faint end has been tested against different lens models, completeness corrections and redshift contamination can still introduce significant uncertainties. Independent verification with other lensing fields will therefore be valuable for establishing the faint-end slope more robustly. When comparing our extrapolated model predictions with $z\gtrsim12$ UVLF observations, it should be noted that the constraints are primarily based on photometric candidates whose observed abundance is subject to redshift uncertainties (including low-$z$ interlopers), completeness corrections, and cosmic variance. 

On the modeling side, our adopted $\sigma_{\rm UV}$ should be interpreted as an effective scatter that in principle includes contributions not only from bursty SFHs, but also from dust attenuation, metallicity, and IMF variations, among other sources of halo-to-halo $M_{\rm UV}$ variations. For a given $\sigma_{\rm UV}$ prescription, the inferred median SFE relation also depends on the adopted parameterization of the target UVLF and numerical implementation and treatment of abundance matching. From our sensitivity tests, variations in $M_{\rm h,min}$ by factors of order unity preserve the qualitative trends, although precise SFE normalizations remain sensitive to modeling and numerical assumptions, particularly at halo masses not far above $M_{\rm h,min}$. In addition, separately calibrated models with large constant scatter can also produce comparable high-$z$ UVLFs. Among our examined $\sigma_{\rm UV}$ prescriptions, the steeper low-mass SFE inferred is a more distinctive outcome of mass-dependent UV variability.

Finally, our extrapolation treats both $\sigma_{\rm UV}$ and the median SFE as redshift-independent functions of halo mass. This represents a simplified baseline assumption motivated by theoretical studies of galaxy formation at cosmic dawn \citep{Ma2018,Feldmann2025,Sun2026}, but its validity across the full mass and redshift range of early galaxies remains to be tested. Density-dependent changes in star formation and feedback could modify the SFE \citep{Dekel2023,Somerville2025}, whereas UV variability at a given mass can also evolve. Improved joint measurements of UVLFs, clustering, and stellar populations by e.g., JWST and Roman will eventually test and refine these assumptions and solidify the basis for interpreting bursty star formation in early galaxies. \\

\begin{figure*}[!ht]
    \centering
    \includegraphics[width=0.875\textwidth]{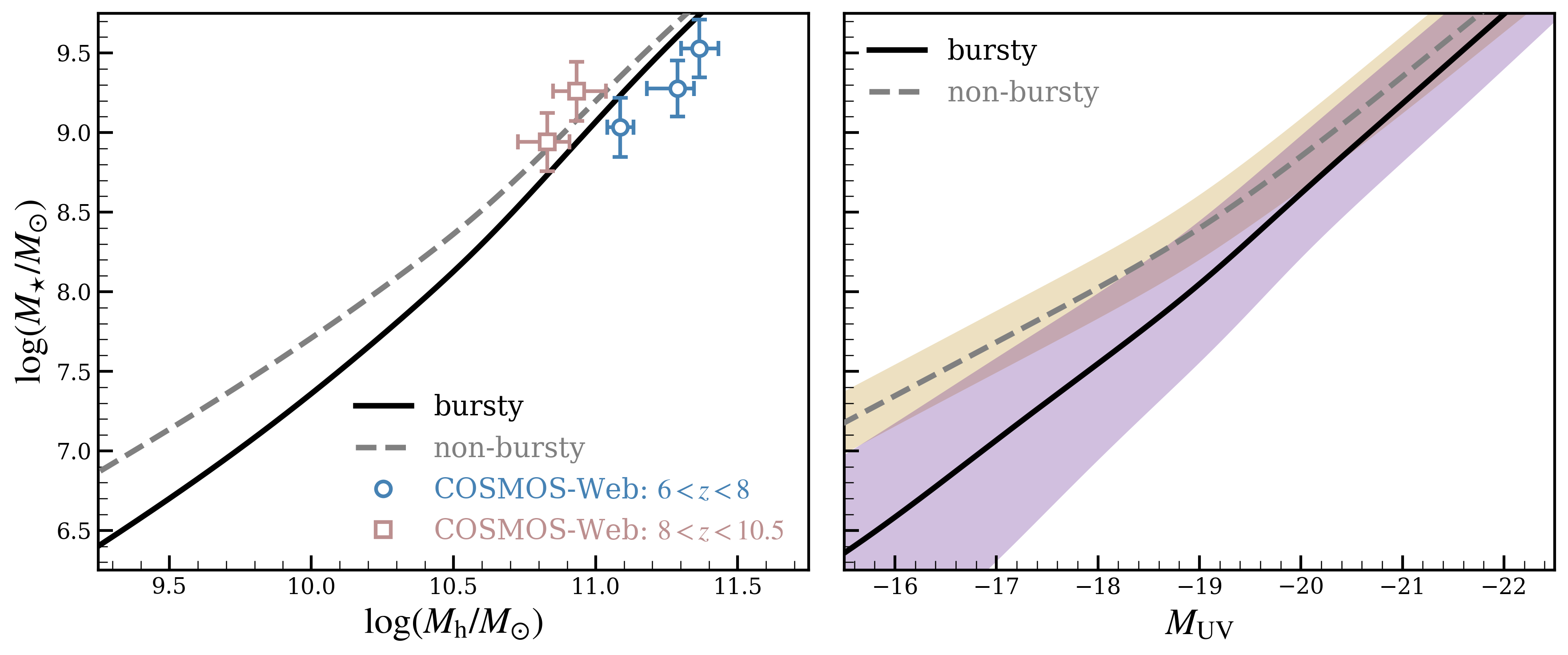}
    \caption{Left: Stellar mass--halo mass relations obtained by integrating the inferred median SFE according to Equation~(\ref{eq:mstar}), compared with clustering-based estimates from the COSMOS-Web survey \citep{Paquereau2025}. The model tracks are redshift-independent by construction. Right: Medians and 16--84th percentiles (shaded regions) of stellar mass at fixed UV magnitude at $z=7$, predicted by the \emph{bursty} (solid) and \emph{non-bursty} (dashed) models. The offset between the medians and the widths of the distributions provide useful diagnostics for examining the burstiness scenarios.}
    \label{fig:mstar}
\end{figure*}

\section*{Acknowledgments}

GS acknowledges support from a CIERA Postdoctoral Fellowship. CAFG was supported by NSF through grants AST-2108230, AST-2307327 and AST-2606015; by NASA through grants 80NSSC22K0809, 80NSSC22K1124 and 80NSSC24K1224; by STScI through grant JWST-AR-03252.001-A; and by BSF through grant \#2024262. SRF was supported by NSF through award AST-2510939. AL acknowledges support from NASA grant 80NSSC26K0183. This work was performed in part at the Aspen Center for Physics, which is supported by the National Science Foundation grant PHY-2210452. ChatGPT 6 Astra was used to assist with refinement of the analysis code and suggest language improvements to the manuscript. All adopted suggestions were reviewed by the authors and any resulting code changes were extensively tested. The authors take full responsibility for the research results and the final manuscript.

\appendix

\section{Stellar Mass Trends Based on the Inferred Median SFE} \label{sec:appendix}

Since Equation~(\ref{eq:sfe}) describes the instantaneous SFE, we consider a toy model that combines it with halo accretion rates to approximate the median stellar mass evolution, which provides an additional diagnostic for comparing burstiness scenarios. Specifically, we calculate the stellar mass--halo mass (SMHM) relation by integrating the inferred median SFE along the halo growth history
\begin{equation}
M_{\star}(M_\mathrm{h})
= (1-R) \int_{t_0}^{t}f_{\star}[M_\mathrm{h}(t')] f_\mathrm{b} \dot{M}_\mathrm{h}(t') d t' = (1-R) \int_{M_\mathrm{h,0}}^{M_\mathrm{h}}f_{\star}(M'_\mathrm{h}) f_\mathrm{b} d M'_\mathrm{h},
\label{eq:mstar}
\end{equation}
where we adopt $M_\mathrm{h,0}=10^{8}\,M_{\odot}$ and the stellar mass return fraction $R \approx 0.2$ appropriate for young stellar populations in high-$z$ galaxies \citep{MaioPeroux2026}. The resulting SMHM relation is redshift-independent. As shown in the left panel of Figure~\ref{fig:mstar}, despite its simplicity, Equation~(\ref{eq:mstar}) predicts SMHM relations broadly consistent with the clustering-based estimates from COSMOS-Web at the redshifts of interest \citep{Paquereau2025}. 

In the right panel of Figure~\ref{fig:mstar}, we use Equation~(\ref{eq:mstar}) to calculate the stellar mass distribution of $z=7$ galaxies selected by $M_{\rm UV}$ in bins of width 1\,mag. The resulting medians and 16--84th percentiles are shown. Compared with the \emph{non-bursty} case, the \emph{bursty} model predicts lower median stellar masses and broader distributions, especially at fainter $M_{\rm UV}$. These differences motivate comparisons with observed stellar mass distributions, although quantitative tests require taking careful account of uncertainties associated with SED fitting and sample selection.

\bibliography{tf}{}
\bibliographystyle{aasjournalv7}



\end{document}